\documentclass[twocolumn,floatfix,showpacs,showkeys,preprintnumbers,nofootinbib,superscriptaddress]{revtex4}
\usepackage[utf8]{inputenc}
\usepackage[sort&compress]{natbib}
\usepackage{ulem}
\usepackage{bm}
\usepackage{times}
\usepackage{amssymb,amsbsy,amsmath,amsfonts}
\usepackage{graphicx}
\usepackage{float}
\usepackage{color}
\usepackage{morefloats}
\usepackage{rotating}
\usepackage{srcltx}
\usepackage{slashed}
\usepackage{subfigure}
\usepackage{multirow}
\usepackage{verbatim}
\usepackage{hyperref}
\usepackage{tabularx}
\usepackage{color}

\begin{document}

\title{Electromagnetic form factors of $\Lambda$ hyperon in the vector meson dominance model and a possible explanation of the near-threshold enhancement of the $e^+e^- \to \Lambda\bar{\Lambda}$ reaction}

\author{Zhong-Yi Li~\footnote{These authors equally contribute to this work. \label{firstcoauthor}}}
\affiliation{Institute of Modern Physics, Chinese Academy of Sciences, Lanzhou 730000, China}
\affiliation{School of Nuclear Science and Technology, University of Chinese Academy of Sciences, Beijing 101408, China}

\author{An-Xin Dai~\textsuperscript{\ref{firstcoauthor}}}
\affiliation{Institute of Modern Physics, Chinese Academy of Sciences, Lanzhou 730000, China}
\affiliation{School of Nuclear Science and Technology, University of Chinese Academy of Sciences, Beijing 101408, China}

\author{Ju-Jun Xie}~\email{xiejujun@impcas.ac.cn}
\affiliation{Institute of Modern Physics, Chinese Academy of Sciences, Lanzhou 730000, China}
\affiliation{School of Nuclear Science and Technology, University of Chinese Academy of Sciences, Beijing 101408, China}
\affiliation{School of Physics and Microelectronics, Zhengzhou University, Zhengzhou, Henan 450001, China}
\affiliation{Lanzhou Center for Theoretical Physics, Key Laboratory of Theoretical Physics of Gansu Province, Lanzhou University, Lanzhou, Gansu 730000, China}

\date{\today}
\begin{abstract}

The near-threshold $e^+e^- \to \Lambda\bar{\Lambda}$ reaction is studied with the assumption that the production mechanism is due to a near-$\Lambda \bar{\Lambda}$-threshold bound state. The cross section of $e^+e^- \to \Lambda\bar{\Lambda}$ reaction is parametrized in terms of the electromagnetic form factors of $\Lambda$ hyperon, which are obtained with the vector meson dominance model. It is shown that the contribution to the $e^+e^- \to \Lambda\bar{\Lambda}$ reaction from a new narrow state with quantum numbers $J^{PC}=1^{--}$ is dominant for energies very close
to threshold. The mass of this new state is around 2231 MeV, which is very close to the mass threshold of $\Lambda \bar{\Lambda}$, while its width is just a few MeV. This gives a possible solution to the problem that all previous calculations seriously underestimate the near-threshold total cross section of the $e^+e^- \to \Lambda\bar{\Lambda}$ reaction. We also note that the near-threshold enhancement can be also reproduced by including these well established vector resonances $\omega(1420)$, $\omega(1650)$, $\phi(1680)$, or $\phi(2170)$ with a Flatt${\rm\acute{e}}$ form for their total decay width, and a strong coupling to the $\Lambda\bar{\Lambda}$ channel.

\end{abstract}
\maketitle

\section{Introduction}

Electromagnetic form factors (EMFFs) is an important tool for studying the electromagnetic structure of hadrons~\cite{Pacetti:2014jai}. The measurements of space-like region EMFFs of proton can be done in elastic as well as inelastic $ep$ scattering~\cite{JeffersonLabHallA:2001qqe}. While for $\Lambda$ hyperon, its EMFFs at the space-like region are very hardly been experimentally measured. Instead, the electron-positron annihilation process, $e^+ e^- \to \Lambda \bar{\Lambda}$ allows to study the $\Lambda$ hyperon EMFFS at the time-like region~\cite{BaBar:2007fsu,BESIII:2017hyw,BESIII:2019nep,Haidenbauer:1991kt,Baldini:2007qg,Faldt:2016qee,Haidenbauer:2016won,Faldt:2017kgy,Yang:2017hao}.
In addition, the $e^+ e^- \to \Lambda \bar{\Lambda}$ reaction can be used to study vector mesons with light quark flavors and mass above 2 GeV~\cite{Bystritskiy:2021frx,Wang:2021gle}, especially for the $\phi$ excitations~\cite{Xiao:2019qhl,Cao:2018kos}.

Recently, the BESIII collaboration measured the $e^{+}e^{-} \to \Lambda \bar{\Lambda}$ reaction with much improved precision. The Born cross-section at the center of mass energy $\sqrt{s}$=2.2324 GeV is determined to be $305 \pm 45^{+66}_{-36}$ pb~\cite{BESIII:2017hyw}. This indicates there is an evident threshold enhancement for the $e^{+}e^{-} \to \Lambda \bar{\Lambda}$ reaction. The observed value is larger than the previous theoretical
predictions, which predicted that the total cross section of the $e^{+}e^{-} \to \Lambda \bar{\Lambda}$ reaction should be close to zero near the reaction threshold. In fact, before the new measurements of Ref.~\cite{BESIII:2017hyw}, there exist several theoretical studies of this reaction, which proposed the final state interactions~\cite{Baldini:2007qg,Haidenbauer:2016won} to explain the unexpected features of the $e^+ e^- \to \Lambda\bar{\Lambda}$ cross sections near threshold.

After the observations of Ref.~\cite{BESIII:2017hyw} by the BESIII collaboration, the $e^{+}e^{-} \to \Lambda \bar{\Lambda}$ reaction was investigated in Ref.~\cite{Cao:2018kos}, where it was found that the $\phi(2170)$ is responsible for the threshold enhancement. In Ref.~\cite{Yang:2019mzq}, by using a modified vector meson dominance (VMD) model, an analysis on the EMFFs of $\Lambda$ hyperon and also the $e^+e^- \to \Lambda\bar{\Lambda}$ reaction was performed, where those contributions from $\phi$, $\omega$, $\omega(1420)$, $\omega(1650)$, $\phi(1680)$, and $\phi(2170)$ were taken into account. In Refs.~\cite{Haidenbauer:2016won,Haidenbauer:2020wyp}, with the role played by the final state interactions of the baryon-anti-baryon pairs, the EMFFs of hyperons ($\Lambda$, $\Sigma$, and $\Xi$) were studied in the timelike region. The threshold enhancement of the $e^+ e^- \to \Lambda\bar{\Lambda}$ reaction was also investigated in Refs.~\cite{Haidenbauer:2016won,Yang:2019mzq,Haidenbauer:2020wyp} from the theoretical side. However, a large finite experimental value on the total cross section of $e^{+}e^{-} \to \Lambda \bar{\Lambda}$ reaction at $\sqrt{s}$=2.2324 GeV cannot be well reproduced~\cite{Haidenbauer:2016won,Yang:2019mzq,Haidenbauer:2020wyp}. Further investigations about the $e^+ e^- \to \Lambda\bar{\Lambda}$ reaction
are mostly welcome.

On the other hand the tails of vectors below threshold have to be detected as large effects in the time-like form factors of $\Lambda$ hyperon and possibly as small structures in $e^+ e^- \to \Lambda\bar{\Lambda}$ reaction near threshold. Indeed, as discussed in Ref.~\cite{Cao:2018kos}, the $\phi(2170)$ plays an important role to reproduce the threshold enhancement. However, the width of $\phi(2170)$, $\Gamma_{\phi(2170)} = 165 \pm 65$
MeV~\cite{ParticleDataGroup:2020ssz}, is too wide, thus it will affect a large energy region. Besides, with the Godfrey-Isgur model, a narrow $\Lambda\bar{\Lambda}$ bound state with quantum numbers $J^{PC} = 1^{--}$ and mass around $2232$ MeV was predicted~\cite{Xiao:2019qhl}, and it has significant couplings to both the $\Lambda\bar{\Lambda}$ and $e^+e^-$ channels. While in Refs.~\cite{Zhao:2013ffn,Zhu:2019ibc}, within the one-boson-exchange potential model, a $\Lambda\bar{\Lambda}$ bound state can be also obtained. This narrow $\Lambda\bar{\Lambda}$ bound state, if really existed, will contribute to the threshold enhancement of the $e^+ e^- \to \Lambda\bar{\Lambda}$ reaction and also the EMFFs of the $\Lambda$ hyperon in the time like region.

In this work we take the achievement of the vector meson dominance model and predictions of the narrow $\Lambda\bar{\Lambda}$ bound state as motivation to explore the electromagnetic form factors of the $\Lambda$ hyperon in the time like region. The EMFFs of baryons have been studied with the VMD model for the proton~\cite{Iachello:1972nu,Iachello:2004aq,Bijker:2004yu,Bijker:2006id}, $\Lambda$ hyperon~\cite{Yang:2019mzq}, $\Sigma$
hyperon~\cite{Li:2020lsb}, and charmed $\Lambda^+_c$ baryon~\cite{Wan:2021ncg}. Following Ref.~\cite{Yang:2019mzq}, we revisit the EMFFs of $\Lambda$ hyperon and the $e^+e^- \to \Lambda\bar{\Lambda}$ reaction near threshold by using the modified VMD model. In addition to the contributions from the ground $\omega$ and $\phi$ mesons, we consider also a new narrow vector meson with mass around $2232$ MeV, as predicted in Ref.~\cite{Xiao:2019qhl}.
Yet, the $\phi(2170)$ resonance is not taken into account in the present work,~\footnote{The contributions from $\omega(1420)$, $\omega(1650)$, and $\phi(1680)$ resonances are not considered either, since their masses are far from the reaction threshold of $e^+e^-\to\Lambda\bar{\Lambda}$, and their contributions could be absorbed into the ground $\omega$ and $\phi$ mesons. Besides, we refrain from including such contributions in this work because the model already contains a large number of free parameters.} since the experimental information of it is still diverse, and the measured mass and width of $\phi(2170)$ resonance are controversial~\cite{BESIII:2018ldc,BESIII:2019ebn,BESIII:2020gnc,BESIII:2020kpr,BESIII:2020vtu,BESIII:2020xmw,Huang:2020ocn,BESIII:2021bjn,BESIII:2021yam}.
Indeed, there have also been different theoretical explanations for $\phi(2170)$
resonance~\cite{Page:1998gz,Barnes:2002mu,Ding:2006ya,Wang:2006ri,Ding:2007pc,Chen:2008ej,MartinezTorres:2008gy,Coito:2009na,Ali:2011qi,Dong:2017rmg,Ke:2018evd,Agaev:2019coa,Li:2020xzs,Malabarba:2020grf,Zhao:2019syt}.

This article is organized as follows: the theoretical formalism of the $\Lambda$ hyperon in the VMD model are shown in the following section. In Sec.~III, we present our numerical results and discussions of the $e^+ e^- \to \Lambda\bar{\Lambda}$ reaction. A short summary is given in the last section.

\section{Theoretical formalism}

In this section, we will briefly review the vector meson dominance model to study the electromagnetic form factors of baryons with spin-$1/2$, and the total cross sections of $e^+e^- \to \Lambda\bar{\Lambda}$ reaction and the effective form factor of $\Lambda$.

\subsection{The vector meson dominance model}

Following Ref.~\cite{Yang:2019mzq}, the electromagnetic current of $\Lambda$ hyperon with spin-$1/2$ in terms of the Dirac form factors $F_{1}(Q^{2})$ and Pauli form factors $F_{2}(Q^{2})$ can be written as
\begin{equation}
J^{\mu}=\gamma^{u}F_{1}(Q^{2})+ {\rm i}\frac{\sigma^{\mu
\nu}q_{\nu}}{2m_{\Lambda}}F_ {2}(Q^2),
\end{equation}
where $F_1$ and $F_2$ are functions of the squared momentum transfer
$Q^2 = -q^2$. In the space like region, $q^2 < 0$, while in the time
like region, $q^2 > 0$. The observed electric and magnetic form
factors $G_{E}(Q^2)$ and $G_{M}(Q^2)$ can be expressed in terms of
Dirac and Pauli form factors $F_1(Q^2)$ and $F_2(Q^2)$ by,
\begin{eqnarray}
    G_E (Q^2)\! &=&\!F_1 (Q^2) - \frac{Q^2}{4M^2_\Lambda} F_2 (Q^2), \label{eq:ge} \\
    G_M (Q^2)\! &=&\!F_1 (Q^2) + F_2 (Q^2) \label{eq:gm}.
\end{eqnarray}

In the VMD model, the virtual photon couples to $\Lambda$ hyperon
through vector mesons, thus the Dirac and Pauli form factors are
parametrized as following~\footnote{There is no contributions from
the $\rho$ meson with isospin $I=1$, because the isospin of
$\Lambda$ hyperon is zero.},
\begin{eqnarray}
F_{1}(Q^{2}) &=& g(Q^{2}) [-\beta_{\omega}-\beta_{\phi} -\beta_x + \beta_{\omega}\frac{m_{\omega}^{2}}{m_{\omega}^{2}+Q^{2}} \nonumber \\
        && +\beta_{\phi} \frac{m_{\phi}^{2}}{m_{\phi}^{2}+Q^{2}} +\beta_{x} \frac{m_{x}^{2}}{m_{x}^{2}+Q^{2}} ], \label{eq:f1} \\
F_{2}(Q^{2}) &=& g(Q^{2})
        [(\mu_{\Lambda}-\alpha_{\phi} -\alpha_x) \frac{m_{\omega}^{2}}{m_{\omega}^{2}+Q^{2}} \nonumber \\
        && +\alpha_{\phi} \frac{m_{\phi}^{2}}{m_{\phi}^{2}+Q^{2}} +\alpha_{x} \frac{m_{x}^{2}}{m_{x}^{2}+Q^{2}}],
        \label{eq:f2}
\end{eqnarray}
with $\mu_{\Lambda}=-0.723 \hat{\mu}_{\Lambda}$ in natural unit,
i.e., $\hat{\mu}_{\Lambda} = e/(2M_{\Lambda})$. The $g(Q^2)$ is the
$\Lambda$ intrinsic form factor, and the other terms in
Eqs.~\eqref{eq:f1} and \eqref{eq:f2} are from the vector mesons
($V$) $\omega$, $\phi$, and a new introduced state, which will be discussed in following.

The intrinsic form factor is a dipole $g(Q^2) = 1/(1 + \gamma
Q^2)^2$, which was well used for the proton
case~\cite{Iachello:1972nu,Iachello:2004aq,Bijker:2004yu,Bijker:2006id},
$\Lambda$ case~\cite{Yang:2019mzq}, and $\Sigma$
case~\cite{Li:2020lsb}. In this work, the parameter $\gamma$ in
$g(Q^2)$ and the coefficients $\beta_{\omega}$, $\beta_{\phi}$,
$\beta_{x}$, $\alpha_{\phi}$, $\alpha_x$, $m_x$ and $\Gamma_x$ are
model parameters, which will be determined by fitting them to the
experimental data on the time like electromagnetic form factors of
$\Lambda$ hyperon. The parameters $\beta_{\omega}$, $\beta_{\phi}$,
$\beta_{x}$, $\alpha_x$, and $\alpha_{\phi}$ represent the products
of a $V\gamma$ coupling and a $V \Lambda \Lambda$ coupling, while
$m_x$ and $\Gamma_x$ are mass and total width of the new vector
state included in this work. It is worth to mention that the VMD
model is valid in both space like and time like regions, the model
parameters in both regions are usually considered to be unified,
thus these parameters are real since the EMFFs of baryons in the
space like region are real.

In the time like region we consider also the width of vector mesons
to introduce the complex structure of the electromagnetic form
factors of $\Lambda$ hyperon~\cite{BESIII:2019nep}. For this
purpose, we need to replace
\begin{eqnarray}
g(Q^2) & \to & \frac{1}{(1-\gamma q^2)^2}, \\
\frac{m_{\omega}^{2}}{m_{\omega}^{2}+Q^{2}}  & \to &  \frac{m_{\omega}^{2}}{m_{\omega}^{2}-q^{2}-i m_{\omega} \Gamma_{\omega}}, \\
\frac{m_{\phi}^{2}}{m_{\phi}^{2}+Q^{2}}  & \to & \frac{m_{\phi}^{2}}{m_{\phi}^{2}-q^{2}-i m_{\phi} \Gamma_{\phi}}, \\
\frac{m_{x}^{2}}{m_{x}^{2}+Q^{2}}  & \to &
\frac{m_{x}^{2}}{m_{x}^{2}-q^{2}-i m_{x} \Gamma_{x}},
\end{eqnarray}
where $q^2 = s$ is the invariant mass square of the $e^+ e^- \to
\Lambda\bar{\Lambda}$ reaction. On the other
hand, we take $m_\omega = 782.65$ MeV, $\Gamma_\omega = 8.49$ MeV,
$m_\phi = 1019.461$ MeV, and $\Gamma_\phi = 4.249$ MeV, as quoted in
the particle data group~\cite{ParticleDataGroup:2020ssz}.

\subsection{Total cross sections of $e^+e^- \to \Lambda\bar{\Lambda}$ reaction and the effective form factor of $\Lambda$ hyperon}

Under the one-photon exchange approximation, the total cross section
of $e^+e^- \to \Lambda\bar{\Lambda}$ can be expressed in terms of
the electric and magnetic form factors $G_{E}$ and $G_{M}$ of the
$\Lambda$ hyperon as~\cite{Denig:2012by}
\begin{eqnarray}
\sigma_{e^+e^- \to \Lambda\bar{\Lambda}} \! = \! \frac{4 \pi \alpha^2
\beta}{3 s^2} \left( s \left|G_{M}\left( s \right)\right|^{2} +
2M^2_\Lambda \left|G_{E}\left( s \right)\right|^{2}\right), \label{eq:tcs}
\end{eqnarray}
where $\alpha = e^2/(4\pi) = 1/137.036$ is the fine-structure
constant and $\beta = \sqrt{1-4 M_\Lambda^{2}/s}$ is a phase-space
factor.

The measurement of the total cross section in Eq.~\eqref{eq:tcs} at
a fixed energy allows for determination of the combination of
$|G_E|^2$ and $|G_M|^2$. With precise measurements of the angular
distributions of the $e^+e^- \to \Lambda\bar{\Lambda}$ reaction, a
separate determination of $|G_E|$ and $|G_M|$ is possible. Instead
of a separation between $G_E$ and $G_M$, one can easily obtain the
effective form factor $G_{\rm eff}(s)$ of the $\Lambda$ hyperon from
the total cross section of $e^+ e^- \to \Lambda \bar{\Lambda}$
annihilation process~\cite{BESIII:2017hyw,BESIII:2019nep}. It is
defined as
\begin{eqnarray}
G_{\mathrm{eff}}\left( s \right) & = & \sqrt{\frac{\sigma_{e^+e^-
\to \Lambda\bar{\Lambda}}}{[1+1/(2\tau)][4\pi\alpha^2\beta/(3s)]}}
\nonumber \\
& = & \sqrt{\frac{2
\tau\left|G_{M}\left(q^{2}\right)\right|^{2}+\left|G_{E}\left(q^{2}\right)\right|^{2}}{1+2
\tau}},
\end{eqnarray}
where $\tau = s/(4 M^2_\Lambda)$. The effective form factor square $G^2_{\rm
eff}(s)$ is a linear combination of $|G_E|^2$ and $|G_M|^2$, and
proportional to the square root of the total cross section of $e^+
e^- \to \Lambda \bar{\Lambda}$ reaction, which is definitely real.
On the other hand, the effective form factor $G_{\rm eff}(s)$
indicates also how much the experimental $e^+ e^- \to \Lambda
\bar{\Lambda}$ cross section differs from a point like $\Lambda$
hyperon.

\section{Numerical results and discussions}

In this work, following Ref.~\cite{Iachello:2004aq}, we will consider only the $\beta_x$ term in the Dirac form factor $F_1$, and $\alpha_x$ is taken as zero.~\footnote{In fact, we found that only the sum of the two parameters $\beta_x$ and $\alpha_x$ can be determined by the $\Lambda$ effective form factor in the energy region of the mass threshold of $\Lambda\bar{\Lambda}$, thus we keep only the $\beta_x$ term, that is $\alpha_x =0$, since $\alpha_x$ is associated to the tensor coupling in Pauli form factor $F_2$.} Then, we perform seven-parameter ($\gamma$, $\beta_{\omega}$, $\beta_{\phi}$, $\beta_{x}$, $\alpha_\phi$, $m_x$ and $\Gamma_x$) $\chi^2$ fits to the experimental data on the effective
form factor $G_{\rm eff}$ of $\Lambda$ hyperon and the form factor
ratio $R = |G_E/G_M|$. There are a total of 18 data points. These
data correspond to the center of mass energy $\sqrt{s}$ ranging from
3.08 down to 2.2324 GeV. The fitted parameters are compiled in
Table~\ref{tab:fittedparameters}, with a reasonably small
$\chi^2/{\rm dof}= 0.9$.

\begin{table}[htbp]
    \caption{\label{tab:fittedparameters} Values of model parameters determined in this work.}
    \begin{ruledtabular}
        \begin{tabular}{cccc}
            \textrm{Parameter}&
            \textrm{Value}&
            \textrm{Parameter}&
            \textrm{Value}\\
            \colrule
             $\gamma$ $({\rm GeV}^{-2})$  & $0.43 ~(0.48 \pm 0.08)$ & $\beta_{\phi}$ & $1.35$  \\
             $\beta_\omega (\beta_{\omega\phi})$  & $-1.13~(-0.21 \pm 0.14)$  & $\alpha_\phi$  & $-0.40$ \\
              $\beta_x$ $(10^{-3})$  & $1.50~(-3.21 \pm 4.35)$  & $\Gamma_{x}(\rm MeV)$ & $4.7 ~ (4.8 \pm 15.6)$ \\
              $m_x$ $({\rm MeV})$ & $2230.9 ~ (2229.1 \pm 11.5)$ & & \\
        \end{tabular}
    \end{ruledtabular}
\end{table}

In the present work, since both the $\omega$ and $\phi$ are far from the mass threshold of $\Lambda\bar{\Lambda}$, the behaviour of the contributions from them are similar, we have performed a new fit, which only the $\omega$ term was considered. Thus, we have assumed that $\beta_\phi = \alpha_\phi = 0$ and $\beta_{\omega\phi} = \beta_\omega$, $m_\omega \to (m_\omega+m_\phi)/2$ and $\Gamma_\omega \to (\Gamma_\omega + \Gamma_\phi)/2$. The fitted parameters are shown in brackets in Table~\ref{tab:fittedparameters}. In the case, we get errors for the fitted model parameters, however, the errors obtained from the fit are large.

\begin{figure}[htbp]
\centering
\includegraphics[scale=0.55]{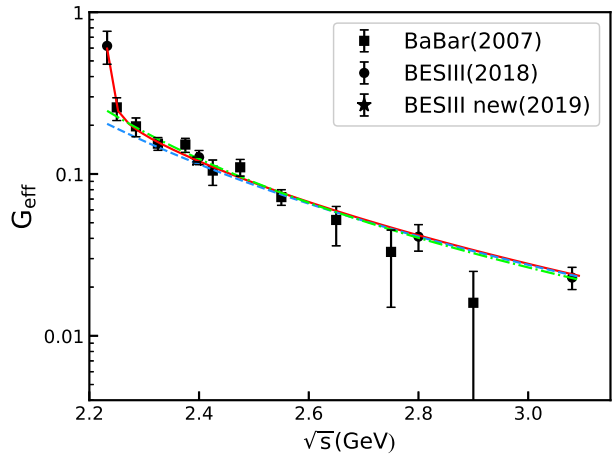}
\caption{The effective form factor $G_{\rm eff}$ of $\Lambda$ hyperon compared with experimental data taken from Refs.~\cite{BaBar:2007fsu,BESIII:2017hyw,BESIII:2019nep}. The red solid curve represents the total contributions from $\omega$, $\phi$ and $X(2231)$, while the blue dashed curve stands for the results without the contribution from the new $X(2231)$ state. The green-dash-dotted curve stands for the fitted results with the effective form factor as in Eq.~\eqref{eq:simpleFF}.} \label{fig:geff}
\end{figure}

In Fig.~\ref{fig:geff} we depict effective form factor $G_{\rm eff}$ of the $\Lambda$ hyperon obtained with the fitted parameters given in
Table~\ref{tab:fittedparameters} of the seven-parameter fit, as a function of $\sqrt{s}$. The experimental data points are taken from Refs.~\cite{BaBar:2007fsu,BESIII:2017hyw,BESIII:2019nep}. The red curve is the total contribution, while the blue dashed curve is the contributions from only $\omega$ and $\phi$, with $\beta_x = 0$. One can see that the experimental data can be well described with the contribution from the new narrow state, especially for the first four data points close to reaction threshold.

Note that the only fifteen available data points about the effective form factor and three data points about the ratio of $R = |G_E/G_M|$ do not allow to obtain unique values for the model parameters, which we introduced for the Dirac and Pauli form factors. Above $\sqrt{s}= 2.3$ GeV, the line shape of $G_{\rm eff}$ is trivial and there are many solutions to describe it. Thus, it is very difficult to get the parameter errors in the fit. In fact, one can also get a good fit to the effective form factor data except the first one with the following parametrized of $G_{\rm eff}$~\cite{Bianconi:2015owa,BESIII:2021dfy}:
\begin{eqnarray}
G_{\rm eff} = C_0 g(q^2) = \frac{C_0}{(1-\gamma q^2)^2}. \label{eq:simpleFF}
\end{eqnarray}
The fitted parameters are $\gamma = 0.33 \pm 0.03$ ${\rm GeV}^{-2}$ and $C_0 = 0.10 \pm 0.03$. The fitted $G_{\rm eff}$ are shown in Fig.~\ref{fig:geff} with the green-dash-dotted curve with central values of $\gamma$ and $C_0$. One can see that the experimental data can be well reproduced except the first point.

To get more precise information of $m_x$ and $\Gamma_x$ obtained from the seven-parameter fit, by fixing other parameters with their values as shown in Table~\ref{tab:fittedparameters}, within the range of $m_x(1\pm10\%)$ and $(0,2\Gamma_x)$, we generate random sets of the fitted parameters $(m_x, \Gamma_x)$ with a Gaussian distribution. For each set of $(m_x,\Gamma_x)$, we perform a $\chi^2$ fit to the first four data points of the effective form factor. We collect these sets of the fitted parameters, such that the corresponding $\chi^2$ are below $\chi^2_{\rm min}+1$, where $\chi^2_{\rm min}$ is obtained with these parameters shown in Table~\ref{tab:fittedparameters}. With these collected best fitted parameters, we obtain the errors of parameters $m_x$ and $\Gamma_x$, which are: $m_x=2230.9^{+3.4}_{-3.5}$ MeV, and $\Gamma_x=4.7^{+2.2}_{-4.7}$ MeV.

One might think that a Flatt${\rm\acute{e}}$ type~\cite{Flatte:1976xu} for $\Gamma_x$ might improve the fitting situation, since the mass of the new state is very close to the $\Lambda\bar{\Lambda}$ threshold, and it may also couples strongly the $\Lambda\bar{\Lambda}$ channel. The Flatt${\rm\acute{e}}$ form is useful for coupled-channel analysis, however, we currently have experimental information about only the $\Lambda\bar{\Lambda}$ channel. Yet, we have explored such a possibility. We take
\begin{eqnarray}
\Gamma_x = \Gamma_0 + \Gamma_{\Lambda\bar{\Lambda}}(s),
\end{eqnarray}
where $\Gamma_0$ is a constant and it includes the contributions from the other channels, while $\Gamma_{\Lambda\bar{\Lambda}}(s)$ is the contribution from the $\Lambda\bar{\Lambda}$ channel. For example, with $s$-wave coupling~\cite{Zou:2002yy} for the new state with $J^{PC} = 1^{--}$ to the $\Lambda\bar{\Lambda}$ channel, one can get~\footnote{In general, there should be also contributions from $d$-wave.}:
\begin{eqnarray}
\Gamma_{\Lambda\bar{\Lambda}}(s) = \frac{g^2_{\Lambda\bar{\Lambda}}}{4 \pi} \sqrt{s/4 - M^2_{\Lambda}},
\end{eqnarray}
where $g_{\Lambda\bar{\Lambda}}$ is the unknown $s$-wave coupling constant.

Then we have performed six-parameter ($\gamma$, $\beta_{\omega\phi}$, $\beta_x$, $m_x$, $\Gamma_0$ and $g_{\Lambda\bar{\Lambda}}$) $\chi^2$ fits. Indeed, we can also obtain a good fit. The fitted parameters are: $\gamma = 0.57 \pm 0.21$ ${\rm GeV}^{-2}$, $\beta_{\omega\phi} = -0.30 \pm 0.31$, $\beta_x = -0.03 \pm 0.09$, $m_x =2237.7 \pm 50.2 $ MeV, $\Gamma_0 = 8.8^{+75.9}_{-8.8}$ MeV, and $g_{\Lambda\bar{\Lambda}} = 3.0 \pm 1.9$. One can find that the fitted errors for the model parameters are very large. On the other hand, one can also look for poles for the Breit-Wigner function, $1/(s-m^2_x + i m_x\Gamma_x)$, which is parametrized by the Flatt${\rm\acute{e}}$ form on the complex plane of $\sqrt{s}$. With the above fitted central values of $m_x$, $\Gamma_0$ and $g_{\Lambda\bar{\Lambda}}$, we get a pole at $\sqrt{s} = Z_R = M_R - i \Gamma_R/2 = (2096.2, -9.9)$ MeV.

As discussed before, only a few experimental data which are very close to and above the reaction threshold need the contribution of the new state. In fact, the Flatt${\rm\acute{e}}$ formulae would push down the Breit-Wigner mass, and there will be a clear drop at the $\Lambda\bar{\Lambda}$ threshold if the value of $g_{\Lambda\bar{\Lambda}}$ is large~\cite{Xie:2013wfa}. However, we donot have any information below the $\Lambda\bar{\Lambda}$ mass threshold, which means that the mass and width of the state still cannot be well determined if we choose the Flatt${\rm\acute{e}}$ formulae. Indeed, we can get good fits by including $\omega(1420)$, $\omega(1650)$, $\phi(1680)$, or the $\phi(2170)$ with a Flatt${\rm\acute{e}}$ form and taking their mass $m_x$ and width $\Gamma_0$ as quoted in PDG. On the other hand, it is worth to mention that, for very wide regions for these values of $m_x$ and $\Gamma_0$ one can always get a good fit by adjusting the value of $g_{\Lambda\bar{\Lambda}}$, this is because we donot have information below the $\Lambda\bar{\Lambda}$ mass threshold. In this work since the $\Lambda\bar{\Lambda}$ channel is opened in the considering energy region, we just use a constant total decay width for this new state, in such a way we can also reduce the number of free parameters. This work constitutes a first step in this direction.

\begin{figure}[htbp]
    \centering
    \includegraphics[scale=0.5]{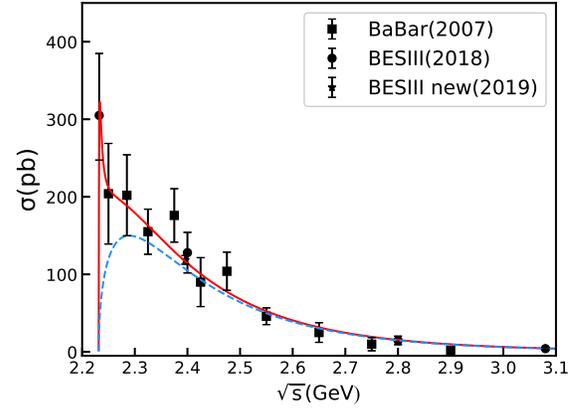}
\caption{The total cross section of $e^+ e^- \to
\Lambda\bar{\Lambda}$ reaction compared with the experimental data
measured by BABAR collaboration~\cite{BaBar:2007fsu} and BESIII
collaboration~\cite{BESIII:2017hyw,BESIII:2019nep}.} \label{fig:tcs}
\end{figure}

Next, we pay attention to the total cross sections of $e^+e^- \to \Lambda \bar{\Lambda}$ reaction. In Fig.~\ref{fig:tcs}, the theoretical fitted results of the total cross sections of the $e^+e^- \to \Lambda\bar{\Lambda}$ reaction in the energy range from the reaction threshold to $\sqrt{s} = 3.1$ GeV are shown and compared to experimental data taken from Refs.~\cite{BESIII:2017hyw,BESIII:2019nep}. In this figure, the red solid line displays the theoretical fitted result with total contributions from $\omega$, $\phi$, and the new state $X(2231)$, while the blue dotted cure represents the results without the contribution from $X(2231)$ state. Again, one can see that the near threshold enhancement structure is well reproduced thanks to a significant contribution from a very narrow state $X(2231)$ with mass about 2231 MeV. The narrow peak of this state is clearly seen.

\begin{figure}[htbp]
    \centering
    \includegraphics[scale=0.5]{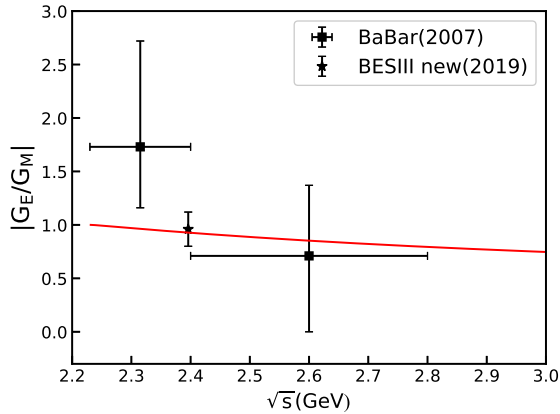}
\caption{Electromagnetic form factor ratio $|G_E/G_M|$ compared with
the experimental data taken from
Refs.~\cite{BaBar:2007fsu,BESIII:2019nep}. }
    \label{fig:ge2gm}
\end{figure}

Finally, in Fig.~\ref{fig:ge2gm} we show the form factor ratio
$|G_E/G_M|$ obtained with the fitted parameters given in
Table~\ref{tab:fittedparameters}, as a function of $\sqrt{s}$. The
experimental data points are taken from
Refs.~\cite{BESIII:2017hyw,BESIII:2019nep}. This ratio is determined
to be one at the threshold due to the kinematical restriction, which
can be easily obtained from Eqs.~\eqref{eq:ge} and \eqref{eq:gm}.

We find that a narrow vector meson, $X(2231)$, whose mass is close
to the mass threshold of $\Lambda\bar{\Lambda}$, is needed to describe the
threshold enhancement of $e^+e^- \to \Lambda\bar{\Lambda}$ reaction.
However, its width cannot be well determined through the VMD model
by fitting the current experimental data. This state could be a quasi-bound-state of $\Lambda \bar{\Lambda}$, which has
significant couplings to the $\Lambda\bar{\Lambda}$ and $e^+e^-$
channels. In fact, from the analysis of the $p\bar{p}\to
\Lambda\bar{\Lambda}$ reaction near threshold, the authors of
Ref.~\cite{Carbonell:1993dt} predicted also a narrow
$\Lambda\bar{\Lambda}$ subthreshold state with quantum numbers
$J^{PC} = 1^{--}$ and has a width of a few MeV. However, a later
high-statistics measurement of the $p\bar{p} \to
\Lambda\bar{\Lambda}$ reaction~\cite{Barnes:2000be} ruled out the
existence of such a $\Lambda\bar{\Lambda}$ resonance as predicted in
Ref.~\cite{Carbonell:1993dt}, since there is no structure in these
new measurements of the $p\bar{p} \to \Lambda\bar{\Lambda}$ reaction
near threshold, and the total cross section is observed to grow
smoothly from threshold with a mix of $S$- and $P$-wave production.

The near threshold region of the $e^+e^- \to \Lambda\bar{\Lambda}$ reaction was also investigated with specific emphasis on the important role played by the $\Lambda\bar{\Lambda}$ final state interaction in Ref.~\cite{Haidenbauer:2016won}, where the $\Lambda\bar{\Lambda}$ potentials were constructed for the analysis of the $p\bar{p} \to \Lambda\bar{\Lambda}$ reaction. The total cross sections reported by the BaBar collaboration~\cite{BaBar:2007fsu} can be well reproduced, but, the new results from the BESIII collaboration~\cite{BESIII:2017hyw} is very difficult to be obtained by the theoretical calculations of Ref.~\cite{Haidenbauer:2016won}. As discussed above, there is no structure in the $p\bar{p} \to \Lambda\bar{\Lambda}$ reaction, and the BESIII results indicate that we do need include such a narrow state which has significant coupling to the $\Lambda\bar{\Lambda}$ channel.~\footnote{To explain the new BESIII results, a very narrow resonance with mass around the $\Lambda\bar{\Lambda}$ was also discussed in Ref.~\cite{Haidenbauer:2016won}.} Yet, such a narrow state may couple weakly to the $p\bar{p}$ channel, thus it was not appear in the $p\bar{p} \to \Lambda\bar{\Lambda}$ reaction~\cite{Barnes:2000be}.

Moreover, it was found that in the processes of $e^+e^- \to
K^+K^-K^+K^-$ and $e^+e^- \to \phi K^+K^-$, the cross sections are
unusually large at $\rm \sqrt{s}=2.2324$ GeV, which indicates that
there should be contributions from a narrow state with mass about
$2232.4$ MeV~\cite{BESIII:2019ebn}. This state could be the vector
meson $X(2231)$ that we proposed here. On the other hand, in the charmed
sector, the $Y(4630)$ and $Y(4660)$ have been studied in the $e^+ e^- \to
\Lambda_c\bar{\Lambda}_c$ reaction by taking into
account also the $\Lambda_c \bar{\Lambda}_c$ final state
interaction~\cite{Guo:2010tk,Lee:2011rka,Cao:2019wwt,Dong:2021juy}.

Finally, one knows that $G_E$ and $G_M$ are complex in the time like
region, and there is a relative phase angle $\Delta\Phi$ between
these two electromagnetic form factors. In addition to the ratio of
$|G_E/G_M|$, a rather large phase $\Delta\Phi = 37^\circ \pm
12^\circ \pm 6^\circ$ was also obtained at $\sqrt{s} = 2.396$ GeV by
the BESIII collaboration~\cite{BESIII:2019nep}. Because the fitted
width of $X(2231)$ is so narrow, we cannot reproduce this large
phase at $\sqrt{s} = 2.396$ GeV. The large phase will be described
by considering these vector mesons with higher masses around $2.3
\sim 2.4$ GeV and wide widths as predicted in
Refs.~\cite{Wang:2021gle,Cao:2018kos}. Indeed, a broad vector meson
with mass of around $2.34$ GeV was introduced to explain the energy
dependent behavior of the cross sections of the $e^+e^- \to
\Lambda\bar{\Lambda}$ reaction above threshold. Clearly, a further
improved investigations needs to consider the contributions from
these higher mass resonances. But, including such contributions, the
electromagnetic form factors of $\Lambda$ hyperon would become more
complex due to additional parameters from the vector meson dominance
model, and we cannot determine or constrain these parameters. In the
present work, we focus on the near threshold enhancement of the
$e^+e^- \to \Lambda\bar{\Lambda}$ reaction. Thus, we will leave
these contributions to future studies when more precise experimental
data become available.

\section{Summary}

In this work, we have studied the $e^+e^- \to \Lambda\bar{\Lambda}$
reaction near threshold and the electromagnetic form factors of the
$\Lambda$ hyperon within the modified vector meson dominance model.
In addition to these contributions from ground $\omega$ and $\phi$
meson, we introduce also a new narrow vector meson $X(2231)$ with
mass around the mass threshold of $\Lambda\bar{\Lambda}$, and its
width is about few MeV. It is found that we can describe the
effective form factor $\rm G_{eff}$ and the electromagnetic form
factor ratio $|G_E/G_M|$ of $\Lambda$ hyperon quite well.
Especially, the threshold enhancement of the total cross sections of
the $e^+e^- \to \Lambda\bar{\Lambda}$ reaction at $\sqrt{s}$=2.2324
GeV can be well reproduced. This narrow state could be a $\Lambda\bar{\Lambda}$ quasi-bound-state with quantum numbers
$J^{PC} = 1^{--}$. Further data in the very close to threshold region with better mass resolution would be very useful to confirm this narrow resonance.

On the other hand, if one take a Flatt${\rm\acute{e}}$ form for the total decay width of $\omega(1420)$, $\omega(1650)$, $\phi(1680)$, and $\phi(2170)$, the experimental data can be also well reproduced with a strong coupling of these resonance to the $\Lambda\bar{\Lambda}$ channel.

The proposed formalism and conclusion here would give insight into
the electromagnetic form factors of the $\Lambda$ hyperon and the
near threshold enhancement of the $e^+e^- \to \Lambda\bar{\Lambda}$
reaction. The proposed formalism attribute the $e^+e^- \to
\Lambda\bar{\Lambda}$ non-vanishing cross sections near threshold to the contribution of a new narrow vector meson $X(2231)$, which
could be the peak structure seen in the $e^+e^- \to K^+K^-K^+K^-$
and $e^+e^- \to \phi K^+K^-$ reactions at $\sqrt{s}=2.2324$ GeV. It
is expected that this conclusion can be distinguished and may be
tested by the future experiments with improved precision at BESIII
or the planned Super tau-charm Facility at
China~\cite{Shi:2020nrf,Sang:2020ksa,Fan:2021mwp}.

%%%%%%%%%%%%%%%%%%%%%%%%%%%%%%%%%%%%%%%%%%%%%%%%%%%%%%%%%%%%%%%%%%%%%%%%%%%%
%
\begin{acknowledgments}

We would like to thank Profs. De-Xu Lin and Hai-Qing Zhou for useful
discussions. This work is partly supported by the National Natural
Science Foundation of China under Grant Nos. 12075288, 11735003, and
11961141012. It is also supported by the Youth Innovation Promotion
Association CAS.

\end{acknowledgments}

\end{document}